\documentclass[aps,pra,twocolumn,floatfix,showpacs]{revtex4-1}
\usepackage{graphicx,amsfonts}
\usepackage{amsmath,amscd,amsfonts,amssymb,color}
\newcommand{\bra}[1]{\left\langle{#1}\right\vert}
\newcommand{\ket}[1]{\left\vert{#1}\right\rangle} 

\begin{document}

\title{Delayed choice of paths selected by grin and snarl of quantum Cheshire Cat}
\author{Debmalya Das\(^{1,2,3}\) and Ujjwal Sen\(^3\)}

\affiliation{
\(^1\)Department of Physics and Astronomy, University of Rochester, Rochester, New York 14627, USA\\
\(^2\)Center for Coherence and Quantum Optics, University of Rochester, Rochester, New York 14627, USA\\
\(^3\)
Harish-Chandra Research Institute,
HBNI, Chhatnag Road, Jhunsi, Prayagraj (Allahabad) 211 019, India
}

\begin{abstract}

The so-called quantum Cheshire Cat is a scenario where a photon, identified with a cat, and a component of its polarization, identified with the grin of that cat, are separated. We observe that the same techniques can be used to separate two orthogonal components of polarization of a photon, on an average. We identify these polarization components of the photon as the grin and snarl of the cat for ease of comprehension. A  gedanken experiment is presented in which we simultaneously tune the input polarizations of the photon in the two arms of a Mach-Zehnder interferometer. It is noted that for two particular choices of photon polarization, the presence of the two components gets reversed in the two arms. This reversal of the grin and the snarl occurs before the polarization components even interact with the tuners, i.e., before the choice of which arm each should be in is made.
\end{abstract}

\maketitle

\section{Introduction}
Experience in the everyday world leads us to believe that an entity and its property are inseparable. In classical
physics this easily plays out. For example, while studying the motion of a rolling ball, we can define physical 
quantities such as velocity, angular velocity or momentum and so on. Yet, all such physical quantities seem to be tied
to the physical object or entity, the ball. It is not possible 
to talk of the velocity, independent of the ball. 
The velocity ``belongs" to the ball. It would be natural to believe that quantum systems and their properties would 
be no different, and so this idea of a physical property belonging to an object was naturally transferred to the worldview espoused by quantum mechanics. 
In 2012, Aharonov \emph{et al.}~\cite{Cheshire} proposed a thought experiment called quantum Cheshire Cat to demonstrate that
surprisingly, in the realm of quantum mechanics,
a particle can be decoupled from its property, under certain conditions. Using a setup based on the Mach-Zehnder
interferometer and a single photon, it can be shown that for a certain combination of preselected and 
postselected states of the photon, it can be made to traverse one arm of the interferometer, while its 
polarization traverses the other arm~\cite{Cheshire}. This means that the polarization, a property of
the photon, can exist independently of the photon itself. Clearly, this is not what we experience in the classical
domain. The phenomenon was called ``quantum Cheshire Cat'', inspired by a character, from the famous novel, Alice in Wonderland~\cite{Carroll1865}, who can make his body disappear, starting with his tail and ending with his grin, so that the grin can exist regardless of the body.

The quantum Cheshire Cat has been observed experimentally using neutron interferometry~\cite{Denkmayr2014} as well as 
photon interferometry~\cite{Correa2015, Sponar2016, Ashby2016}. A recent work deals with the phenomenon in the presence
of decoherence~\cite{Richter2018}.
A refined version of the original proposal has been suggested in Ref.~\cite{CC} that decouples all the components 
of the polarization from the photon. Another proposal that deals with the separation of two degrees of freedom
belonging to the same photon can be found in Ref.~\cite{Twin}. Some of the recent works in this area include the 
teleportation of the decoupled circular polarization without the photon~\cite{Das2019a} and exchanging the decoupled
circular polarizations of two quantum Cheshire photons~\cite{Das2020}. An interesting case of the quantum Cheshire Cat
arising from the three-box paradox can be found in Ref.~\cite{Pan2013}. Further discussions on the topic can be found
in Refs.~\cite{Duprey2018, Atherton2015}.

It is to be noted that, in the quantum Cheshire Cat setup, to detect the 
photon
or the polarization in an arm of the interferometer, it is necessary to perform weak measurements and not
strong measurements, as the latter would destroy the preselected state completely and modify the probability
distribution leading to the postselected state.
In order to extract information about a quantum system, 
without significantly altering it, a technique known as weak value measurement was developed in 
Ref.~\cite{AAV1988}. The quantum system is initially prepared in a pure state $\ket{\Psi_{in}}$, also known 
as the preselected state. The weak measurement is executed by weakly coupling the system and a 
meter. After performing the weak measurement of the observable $A$, a projective
or strong measurement of a second observable $B$, which in general does not commute with $A$, is performed
and one of the outcomes, $\ket{\Psi_f}$ is selected. This process is known as postselection. The 
average of the shift in the meter readings, for the weak measurement, corresponding to the postselected state,
is known as the weak value of the observable $A$. The weak value $A_w$ is interpreted as the value of an observable
$A$, between two strong measurements, one giving rise to the preselected state $\ket{\Psi_{in}}$ and the 
other producing the postselected state $\ket{\Psi_f}$.

The weak value is defined as
\begin{equation}
 A_w=\frac{\bra{\Psi_f}A\ket{\Psi_{in}}}{\bra{\Psi_f}\Psi_{in}\rangle}.
 \label{A_w}
\end{equation}
Although interpreted as a value of an observable, the weak value can lie outside the eigenvalue spectrum
~\cite{AAV1988, Duck1989, Tollaksen2010} and can even be complex~\cite{Jozsa2007} with the imaginary part being 
related to the shift in the momentum of the pointer. Weak values have been experimentally observed in 
Refs.~\cite{Pryde2005, Hosten2008, Lundeen2011, Denkmayr2014,  Correa2015, Cormann2016, Sponar2016, Ashby2016}.

While there has been a great deal of debates and discussions on the meaning and interpretation of weak 
value and on the implications it can have on the foundations of quantum mechanics~\cite{Aharonov1990, Tollaksen2007,
Tollaksen2010}, the concept has also found 
myriad applications, including signal amplification~\cite{Dixon2009}, spin Hall effect~\cite{Hosten2008}, quantum state
tomography~\cite{Hofmann2010, Wu2013}, geometric description of quantum states~\cite{Cormann2017}, state visualization
~\cite{Kobayashi2014}, directly measuring the 
wave function of a photon~\cite{Lundeen2011, Lundeen2012}, measuring the expectation value of non-Hermitian operators
~\cite{Pati2015, Nirala2019}, and quantum thermometry~\cite{Pati2019}. 
Weak values have also led to unearthing the possibilities of 
a number of 
results such as the Hardy
effect~\cite{Dolev2005} and the three-box 
effect~\cite{Tollaksen2010}. The quantum
Cheshire Cat is one such phenomenon~\cite{Cheshire} that can be addressed using the weak value machinery.

In this paper, we unveil yet another 
aspect of the phenomenon, related to quantum Cheshire Cat and based on weak values. Instead of a photon and a 
polarization component, we deal with two different components of the polarization of the photon, which are
known to be detected in two different arms of the Mach-Zehnder interferometer. A mechanism to tune the linear 
polarization of the photon is introduced in the two arms of the interferometer. We show that for two
configurations of this tuner, i.e., for two different linear polarizations, the presence of the two components 
of the linear polarization is interchanged in the two arms (as reflected in the corresponding weak values). The element of surprise in this is the fact that
while the two components have to separate in the two arms at a beam splitter, the phase-shifting tuners 
decide which path each should take, at a different point in space, and later in time.
Yet each component is communicated which arm it should be detected in.
We believe that this is an important aspect in the phenomenon of
the quantum Chesire Cat, and can potentially lead to its better
understanding. It also has the potential of being useful in applications in quantum metrology. See \cite{Chesire-cat-in-metro} in this regard.

The paper is organized as follows. Firstly, in Sec.~\ref{grin-snarl}, we discuss how two components of polarization of a photon can be separated. In Sec.~\ref{delay}, we present our thought experiment that exposes the 
paradox of the splitting of the components of polarization with the ``which-path" decision being made at a different 
location and time. We conclude with some discussion on the implications of this paradox in Sec.~\ref{conc}.


\section{Grin and snarl of a quantum Cheshire Cat}
\label{grin-snarl}
The basic exercise in a quantum Cheshire Cat thought-experiment is to perform measurements weakly and under suitable preselection and postselection in the left (\(\ket{L}\)) and the right (\(\ket{R}\)) arms of a Mach-Zehnder interferometer. The weak values of the operators \(\Pi_L=\ket{L}\bra{L}\), \(\Pi_R=\ket{R}\bra{R}\), \(\sigma_x^L=\Pi_L\otimes \sigma_x\) and \(\sigma_x^R=\Pi_R\otimes \sigma_x\) are sought, with \(\sigma_x=\ket{H}\bra{V}+\ket{V}\bra{H}\). These operators correspond to the measurements of the position of a photon and the position of the \(x\) component of polarization in the two arms. If a photon, starting with a horizontal polarization \(\ket{H}\) is prepared in a state
\begin{equation}
    \ket{\Psi_{in}}=\frac{1}{\sqrt{2}}(i\ket{L}+\ket{R})\ket{H},
\end{equation}
and the postselected state is 
\begin{equation}
 \ket{\Psi_f}=\frac{1}{\sqrt{2}}(\ket{L}\ket{H}-i\ket{R}\ket{V}),
\end{equation}
where \(\ket{V}\) denotes vertical polarization, then the aforesaid weak values are measured to be
\begin{equation}
(\Pi_L)_w=1\;\;\;\;\textrm{and}\;\;\;\; (\Pi_R)_w=0, 
\end{equation}
 and 
\begin{equation}
(\sigma_x^L)_w=0\;\;\;\;\textrm{and}\;\;\;\; (\sigma_x^R)_w=1.
\end{equation}
 These indicate that the photon passed through the left arm while the \(x\) component of its polarization passed through the 
right arm. Thus, the polarization component can exist without the presence of the photon in the right arm, a
situation best described by the analogy with \textit{`a grin without a cat!....'} from Alice in Wonderland
~\cite{Carroll1865}.

So far we have concentrated on decoupling a property (polarization-component/ grin) of a system from the system
(photon/ cat) itself. Let us now look at two different properties of the same system, the \(x\) component of polarization
 and the \(z\) component of the polarization, labeled as the grin and the snarl of the cat, respectively. Notice that
 the presence of 
 these two properties can be detected in the left and right arms of the interferometer by weakly measuring the 
 following observables. For the grin, we must weakly measure
  \begin{equation}
  \sigma_x^L=\Pi_L\otimes \sigma_x\; \;\;\;\textrm{and} \;\;\;\; \sigma_x^R=\Pi_R\otimes \sigma_x.
  \label{xcomp}
 \end{equation}
 
On the other hand, for the snarl, we must perform weak measurements of
\begin{equation}
  \sigma_z^L=\Pi_L\otimes \sigma_z\;\;\;\; \textrm{and} \;\;\;\; \sigma_z^R=\Pi_R\otimes \sigma_z,
  \label{zcomp}
 \end{equation}
 where $\sigma_z=\ket{H}\bra{H}-\ket{V}\bra{V}$.

 Under the same preselection $\ket{\Psi_{in}}$ and the postselection $\ket{\Psi_f}$, the weak values for these
 operators are measured to be
 \begin{equation}
  (\sigma_x^L)^w=0\;\;\;\;\textrm{and}\;\;\;\;(\sigma_x^R)^w=1,
 \end{equation} 
 \begin{equation}
  (\sigma_z^L)^w=1\;\;\;\;\textrm{and}\;\;\;\;(\sigma_z^R)^w=0\nonumber.
  \end{equation}
  
Therefore, while the \(x\) component (the grin) of polarization is detected in the left arm, the \(z\) component (the snarl)
is detected in the right arm. Thus two 
properties ($\sigma_x$ and $\sigma_z$) of the same system have been 
``separated." The operators corresponding to the two properties do not commute. In the succeeding section, we add a fresh twist to the tale by arguing that this separation can be controlled 
non-locally, by preparing the preselected state after the point of separation.

\section{Delayed choice of polarization and its effect on grin and  snarl of the quantum Cheshire Cat}
\label{delay}
\begin{figure}
\begin{center}
 \includegraphics[scale=0.8]{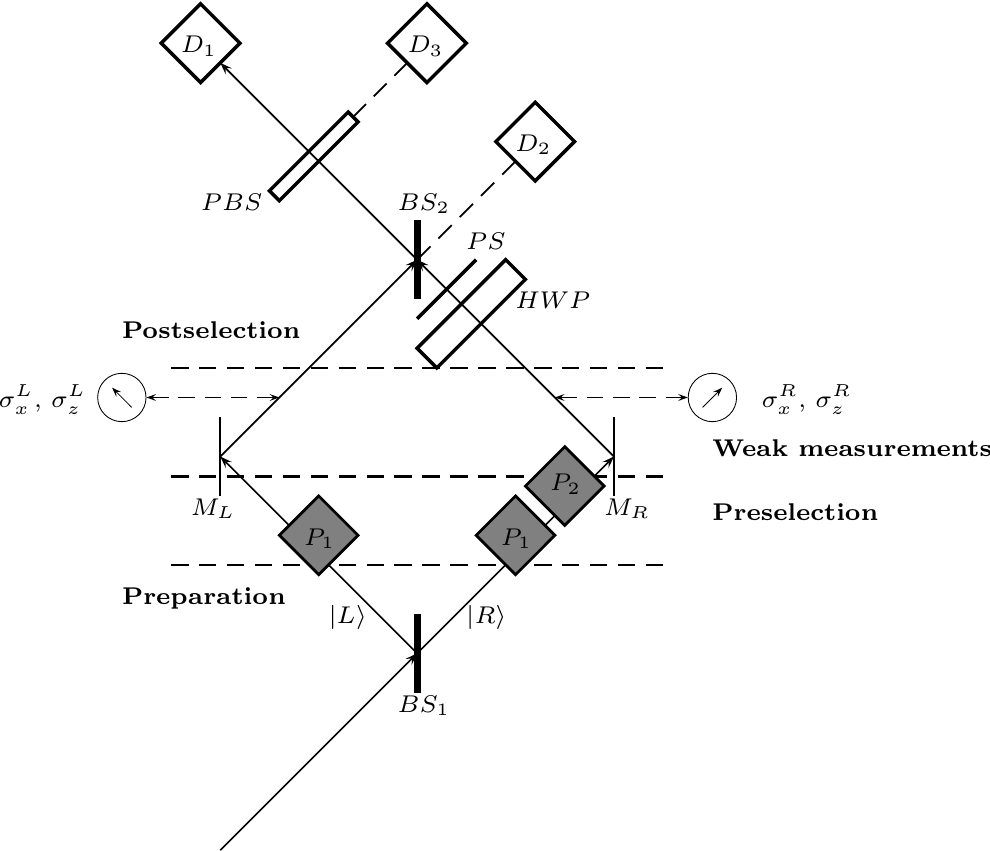}
 \caption{Configuration that decouples grin and snarl of quantum Cheshire Cat with a delayed choice. A set of two 
 phase-tuners $P_1$
 have been introduced in the two arms of the Mach-Zehnder interferometer that causes a rotation of the polarization by
 a phase $\theta$. Another tuner $P_2$ has been included in the right arm that causes a cumulative phase difference of
 $\phi$ in the two paths of the photon. Preselection occurs only after the photons pass through the phase-tuners and 
 along with the choice of postselection, their settings decide which way the $x$ component (grin) and the $z$ component
 (snarl) of the polarization would go. However,
 the separation of the two components must take place from the beam splitter $BS_1$ itself, giving rise to the paradox.}
 \label{DC}
 \end{center}
\end{figure}
We have seen that the preselected state preparation and the decoupling of the grin and the cat or the 
separation of the grin and the snarl, occur at a particular point in space i.e., at the beam-splitter $BS_1$. In this section, 
we relocate the preselection beyond the point in the interferometer where the separation of the two degrees of
freedom occurs. To make this happen, we introduce tunable polarization phase shifters in the two arms of the
interferometer, which are synchronized and can be adjusted simultaneously (see Fig.~\ref{DC}). If we consider that the photon 
entering $BS_1$ has a polarization state $\ket{H}$, then the phase shifters $P_1$ cause a transformation
$\ket{H}\rightarrow \cos \frac{\theta}{2}\ket{H}+ \sin \frac{ \theta}{2}\ket{V}$. The state of the
photon, post these phase-shifters, is given by
\begin{equation}
 \ket{\Psi_{in}^\prime}=\frac{1}{\sqrt{2}}(\ket{L}+e^{i\phi}\ket{R})(\cos \frac{\theta}{2}\ket{H}
 + \sin \frac{\theta}{2}\ket{V}),
 \label{new_pre}
\end{equation}
where we have installed a second adjustable phase shifter $P_2$ in the right arm that causes a total phase difference
of  $\phi$ between the left and the right path degrees of freedom. Note that an extra phase of $\frac{\pi}{2}$
is added to the state associated with the left path, due to reflection from $BS_1$. The phase $\phi$ is the sum
of this phase and the phase difference we would like to introduce using $P_2$. In this thought experiment, we use the state $\ket{\psi_{in}^\prime}$, defined in Eq.(~\ref{new_pre}) as the preselected state.

The detectors in the two arms of the arrangement, measure the observables defined in Eqs.~(\ref{xcomp}) and 
(\ref{zcomp}) which correspond to the presence or absence of two orthogonal components of photon polarizations. These detectors carry out the measurements weakly, which we discuss in more detail later. This is done to ensure that the process of measurement does not alter the probability distribution of the preselected state significantly. However, to obtain weak values, it is also necessary to perform postselection in some desired state. The postselection is carried out in the state
\begin{equation}
 \ket{\Psi_f^\prime}=\frac{1}{\sqrt{2}}(\ket{L}\ket{H}+\ket{R}\ket{V}).
 \label{new_post}
\end{equation}
This is done by a joint arrangement of a half-waveplate (HWP), phase shifter (PS), a second beam-splitter ($BS_2$), a polarization beam splitter (PBS) and three photon detectors $D_1$, $D_2$, and $D_3$ as shown in Fig.~\ref{DC}. The operation of the postselection arrangement is exactly the same as that used in Ref.~\cite{Cheshire}. The clicks of detector $D_1$ gives us the desired state $\ket{\Psi_f^\prime}$. Thus, we have a setup to obtain weak values with $\ket{\Psi_{in}^\prime}$ as the preselected state and $\ket{\Psi_f^\prime}$ as the postselected state. Accordingly, the weak values are given by
\begin{eqnarray}
(\sigma_x^L)_w^\prime &=&\frac{\sin\frac{\theta}{2}}{\cos\frac{\theta}{2}+\sin\frac{\theta}{2}e^{i\phi}},\nonumber\\
(\sigma_x^R)_w^\prime &=& \frac{\cos\frac{\theta}{2}e^{i\phi}}{\cos\frac{\theta}{2}+\sin\frac{\theta}{2}e^{i\phi}}
\label{x}
  \end{eqnarray}  
and
 \begin{eqnarray}
(\sigma_z^L)_w^\prime &=& \frac{\cos\frac{\theta}{2}}{\cos\frac{\theta}{2}+\sin\frac{\theta}{2}e^{i\phi}},\nonumber\\
(\sigma_z^R)_w^\prime &=& -\frac{e^{i\phi}\sin\frac{\theta}{2}}{\cos\frac{\theta}{2}+\sin\frac{\theta}{2}e^{i\phi}}.
\label{z}
  \end{eqnarray}

Suppose that the tuner $P_1$ has been set to $\theta=\pi$. We can immediately see from Eq.~(\ref{z}), that for an arbitrary value of $\phi$,
\begin{equation}
 (\sigma_z^L)_w^\prime=0\;\;\;\;\textrm{and}\;\;\;\;(\sigma_z^R)_w^\prime=-1.
 \label{grin-snarl-.}
\end{equation}
For the same value of $\theta$, and $\phi$ set to $0$, we can see from Eq.~(\ref{x}),
\begin{equation}
 (\sigma_x^L)_w^\prime=1\;\;\;\;\textrm{and}\;\;\;\;(\sigma_x^R)_w^\prime=0.
 \label{grin-snarl-pi}
\end{equation}

In the process of the experiment, one actually needs to measure these values rather than just calculating them from Eq.~(\ref{A_w}). We discuss the measurement process by means of an example, drawing upon the technique used in Ref.~\cite{Das2020}. Suppose we choose to use a qubit system, defined in the basis \(\{\ket{0}_m,\ket{1}_m\}\) as a meter. While measuring, say \(\sigma_z^L\), let us use a joint unitary \(U_{\sigma_z^L}\) to couple the \(z\) component of the photon polarization to the meter. This unitary is given by
\begin{equation}
    U_{\sigma_z^L}=\frac{1}{\sqrt{2}}[(I-\Pi_L)\otimes \sigma_z \otimes I+\Pi_L\otimes\sigma_z\otimes R^{-1}(\theta_g)Z R(\theta_g)].
\end{equation}
Each term consists of a tensor product of operators acting respectively on the photon path, polarization component, and lastly on the meter. The operator \(Z=\ket{0}_m\bra{0}_m-\ket{1}_m\bra{1}_m\) and \(R(\theta_g)\) is defined such that
\begin{eqnarray}
R(\theta_g)\ket{0}_m=\cos{2\theta_g}\ket{0}_m+\sin{2\theta_g}\ket{1}_m,\nonumber\\
R(\theta_g)\ket{1}_m=\sin{2\theta_g}\ket{0}_m-\cos{2\theta_g}\ket{1}_m,
\end{eqnarray}
given \(g=4\theta_g\). The measurement can be made weak by making the coupling between the polarization component and the meter small. This can, in turn, be realized by taking small values of \(g\) so that only terms that are first order in \(g\) can make a significant contribution. Conditioned on the postselection, the meter goes to a state given by
\begin{equation}
    \ket{\Phi}_m=\ket{0}+g(\sigma_z^L)_w\ket{1}.
\end{equation}
Thus, the meter shows a shift that is proportional to the measured weak value.
This weak value turns out to be identical to the one in Eq.~(\ref{grin-snarl-.}). Similarly, joint unitaries can be concocted for the other operators whose weak values need to be measured. For more details and  understanding on the construction of these joint weak unitary evolutions, see~\cite{Kim2018, Pryde2005}.

It is clear that for the above choice of the phases, the \(z\) component of the polarization can be detected only in
the right arm, while simultaneously, the \(x\) component can be detected only in the left arm. Now consider the choice
of the phases as $\theta=0$ and $\phi=0$. The corresponding weak values are
\begin{equation}
 (\sigma_z^L)_w^\prime=1\;\;\;\;\textrm{and}\;\;\;\;(\sigma_z^R)_w^\prime=0,
\end{equation}
\begin{equation}
 (\sigma_x^L)_w^\prime=0\;\;\;\;\textrm{and}\;\;\;\;(\sigma_x^R)_w^\prime=1.
\end{equation}
Clearly, the \(z\) component of polarization can now be detected only in the left arm and the \(x\) component of polarization
can be detected only in the right arm.

The two configurations of the tunable phase shifters can be used to flip the \(x\) and \(z\) components of the polarization of 
the photon between the two arms of the interferometer. What is perplexing is that the components ``have information'' which arm to enter 
even before they encounter the phase shifters $P_1$ and $P_2$ which actually dictate this very occurrence!
It is important to note that the phrase, ``the components have information", is used with the non-quantum assumption that the components of magnetization exist in the photon before the measurement apparatus. This is a typical realist (hidden variable) assumption, and is used to interpret the results and is not used to obtain the results. 
Also, the phrase ``have information'' is used in a non anthropomorphic sense, like the information contained in a quantum state of a physical system.
Thus the grin and
the snarl of 
the quantum Cheshire Cat, not only can travel independently of each other, for the given combination of pre and postselected 
states, they also seem to be affected non locally and from the future, by the configurations of the phase tuners.

One might argue that since weak value is an average shift of the meter, conditioned on the preselected and the postselected states, the two components do not actually travel separately in each arm. Rather they do so only on an average. In our defense, we would like to remind the reader that the original quantum Cheshire Cat is itself, by that logic, an average effect, with the photon and the polarization only decoupling on an average. However, we have taken care to ensure that the weak values observed in this case, are not classical averages but quantum mechanical mean shifts, obtained by considering coupling between the photon or polarization component and a meter.

\section{Conclusion}
\label{conc}
The so-called quantum Cheshire Cat is a phenomenon, in which a property
of a physical system can be temporarily decoupled, i.e. separated, from
the system. The phenomenon provides interesting perspectives in the foundational aspects of
quantum mechanics  and may have future applications. A
corollary of the effect is that two properties, corresponding to
non commuting observables of the same system can be decoupled as well.
Within a related but different setting, we have proposed a gendanken
experiment in which we show that the decoupling and the eventual temporary locations of the properties in separate regions can be affected by an operation at a different location and time. In the spatial trajectory of the properties through the arms of the interferometer, the locations of the tuners performing this operation are reached after the effect takes place. In terms of time of occurrence, the effect is realized even before the operation is performed.

The effect of ``delayed choice" can be realized also by using a photon and its polarization like in quantum Cheshire Cat. However, we have chosen to use two symmetric properties i.e. polarization components in both arms, to demonstrate the effect.

It is interesting to note that weak value is often seen as a result of ``future affecting the past"~\cite{Aharonov2015}. It refers to the choice of postselection, occuring in the future, influencing the meter readings of the measurement centered around the weak value, taking place in the past. The thought experiment, discussed in this paper, describes a different kind of interaction between the future and the past. Here the preselection itself is comprised of three separate events, notably, the preparation of a photon in polarization \(\ket{H}\), passing of the photon through the beam splitter, and finally setting the tuner to the desired configuration. It seems that when coupled to a weak measurement and postselection setup after this, the last stage of the preselection influences the second stage of the same.

Future endeavours in this direction may involve perfecting the understanding of this effect from a foundations point of view. It may also lead to new avenues in designing quantum information protocols.

\begin{acknowledgments}
We acknowledge support from the Department of Science and Technology, Government of India through the QuEST grant
(Grant No. DST/ICPS/QUST/Theme-3/2019/120).
\end{acknowledgments}

\end{document}